\documentclass[aps,jcp,preprint,superscriptaddress,floatfix]{revtex4}

\usepackage[user=AED]{trkchg}    
\usepackage[usera=MM]{trkchgmulti}
\usepackage{color}
\usepackage{dcolumn} 
\usepackage{bm} 
\usepackage{graphicx}
\usepackage{multirow} 
\usepackage{pifont} 
\usepackage{epsfig}
\usepackage{verbatim}
\usepackage{subfigure}
\usepackage{float}
\usepackage{pdfpages}
\usepackage{pdflscape}
\usepackage{amsmath,amsfonts,amssymb,bm,graphicx,color,braket}
\usepackage[version=4]{mhchem}
\usepackage{array}
\usepackage[columnwise]{lineno}
\usepackage{mathrsfs}
\usepackage[nolist]{acronym}
\usepackage[mathcal]{euscript}
\usepackage[titletoc,toc,title]{appendix}
\usepackage{textcomp}
\usepackage{threeparttable}
\usepackage{enumitem}

\newcommand{\ie}{\emph{i.e.}}
\newcommand{\al}{\alpha}
\newcommand{\be}{\beta}

\newcommand{\la}{\lambda}

\newcommand{\ze}{\zeta}
\newcommand{\om}{\omega}

\newcommand{\sig}{\sigma}

\newcommand{\br}{\mathbf{r}}

\newcommand{\EMCPDFT}{E_\text{MCPDFT}}
\newcommand{\ELMCPDFT}{E_\text{$\la$-MCPDFT}}

\newcommand{\MAE}{\overline{\text{MAE}}}
\newcommand{\MAEe}{\text{MAE/e}}
\newcommand{\AKEEe}{\text{AKEE/e}}
\newcommand{\MAKEE}{\overline{\text{MAKEE}}}

\newcommand{\NIAD}{\overline{\text{NIAD}}}

\newcommand{\mc}{\multicolumn}
\newcommand{\mr}{\multirow}

\setenumerate[1]{label=\thesection.\arabic*.}

\begin{document}
	
\author{Mohammad Mostafanejad}
\affiliation{
	Department of Chemistry and Biochemistry,
	Florida State University,
	Tallahassee, FL 32306-4390}
\author{Marcus Dante Liebenthal}
\affiliation{
	Department of Chemistry and Biochemistry,
	Ithaca College,
	Ithaca, NY 14850
}
\affiliation{
	Department of Chemistry and Biochemistry,
	Florida State University,
	Tallahassee, FL 32306-4390}
\author{A. Eugene DePrince III}
\affiliation{
	Department of Chemistry and Biochemistry,
	Florida State University,
	Tallahassee, FL 32306-4390}
\email{deprince@chem.fsu.edu}

\title{Global hybrid multiconfiguration pair-density functional theory}


	
\begin{abstract}

A global hybrid extension of \ac{v2RDM}-driven \ac{MCPDFT} is developed.
Using a linear decomposition of the electron-electron repulsion term, a
fraction $\la$ of the nonlocal exchange interaction, obtained from
\ac{v2RDM}-driven \ac{CASSCF} theory, is combined with its local
counterpart, obtained from an \acl{OTPD} functional. The resulting scheme
(called \acs{lMCPDFT}) inherits the benefits of \ac{MCPDFT} ({\em e.g.},
its simplicity and the resolution of the symmetry dilemma), and, when
combined with the \ac{v2RDM} approach to CASSCF, \acs{lMCPDFT} requires
only polynomially scaling computational effort.  As a result, it can
efficiently describe static and dynamical correlation effects in strongly
correlated systems. The efficacy of the approach is assessed for several
challenging multiconfigurational problems, including the dissociation of
molecular nitrogen, the double dissociation of a water molecule, and the
1,3-dipolar cycloadditions of ozone to ethylene and ozone to acetylene in
the O3ADD6 benchmark set.




\end{abstract}
	
	\maketitle
	
	\section{Introduction}

\label{SEC:INTRODUCTION}

\acresetall



The search for a universally accurate and efficient delineation of
electron correlation effects 
is an active area of research in modern
electronic structure theory. \cite{Helgaker:2000:book,Szabo:1996:book} 
For strongly correlated systems, in particular, where the multiconfigurational character of
the wave function cannot be ignored, the single-determinental form of
Kohn-Sham \ac{DFT} will fail, often dramatically so. In such cases, a
natural remedy is to combine \acl{MR}\acused{MR} and \ac{DFT} methods
(\ac{MR}+\ac{DFT}) to separately model strong and weak correlation effects,
respectively.\cite{Ghosh:2018:7249}
The
development of \ac{MCPDFT}\acused{DFT}\acused{PDFT}
\cite{LiManni:2014:3669} represents an important step in this
direction, as this approach effectively capitalizes on the complementary
strengths of \ac{CASSCF}
theory\cite{Roos:1980:157,Roos:1987:399,Siegbahn:1980:323,Siegbahn:1981:2384}
and \ac{DFT} to offer a robust description of nondynamical correlation
effects and an ecomonical representation of dynamical correlation.
Moreover, \ac{MCPDFT} resolves Kohn-Sham DFT's symmetry dilemma by relying
on functionals of the total density and the \ac{OTPD}
\cite{Becke:1995:147,Gagliardi:2017:66,LiManni:2014:3669} (as opposed to
functionals of the spin-density), and the approach avoids double counting
of electron correlation within the active space.  Recently, we presented
\cite{Mostafanejad:2019:290} a \ac{RDM}-based formulation of \ac{MCPDFT}
that retains these nice properties while significantly reducing the
computational cost of the underlying CASSCF calculations. As such,
variational two-electron RDM (v2RDM)\acused{v2RDM} driven MCPDFT can be
applied to challenging multireference problems that require the
consideration of large active spaces.





Because commonly used\cite{Gagliardi:2017:66,LiManni:2014:3669} \ac{OTPD}
functionals and their conventional Kohn-Sham \ac{DFT} counterparts only
differ only in the input densities (as opposed to their actual functional
forms), the \ac{MCPDFT} suffers from the same ``overbinding tendency'' or
\acl{DE} \cite{Cohen:2008:792} exhibited by familiar \ac{LSDA} and
\ac{GGA} \ac{XC}
functionals.\cite{Becke:1992:2155,Becke:1992:9173,Becke:1993:5648,Becke:1993:1372,Becke:1996:1040}
This overbinding error arises because the electron-gas model
\cite{Giuliani:2005:BOOK} leads to an incorrect picture of the
noninteracting limit, in which the Coulomb repulsion coupling strength is
zero.  The global hybrid scheme offers the simplest remedy for \acl{DE} by
replacing a fraction of the electron-gas exchange with its exact
(Hartree-Fock) counterpart, which then leads to the correct noninteracting
limit.\cite{Becke:1992:2155,Becke:1992:9173,Becke:1993:5648,Becke:1993:1372,Becke:1996:1040}



In light of the great success of the global hybrid scheme in standard
Kohn-Sham \ac{DFT}, several proposals have been offered that extend this
approach to the multiconfiguration
regime\cite{Sharkas:2011:064113,Sharkas:2012:044104,Garza:2015:044109,Garza:2015:22412,Ying:2019:225,Toulouse:2011:101102};
for a survey of these techniques, the reader is referred to
Ref.~\citenum{Ghosh:2018:7249}.  The present study is inspired by the
\ac{MC1H} method of Sharkas {\em et al.}\cite{Sharkas:2012:044104}, which is
similar to \ac{MCPDFT} in that it captures nondynamical correlation with \ac{MCSCF} theory
and the remaining dynamical correlation effects via \ac{DFT}. 
The \ac{MC1H} method has motivated the
development of other \ac{MR}+\ac{DFT} techniques, such as \ac{pCCD-lDFT}
\cite{Garza:2015:044109,Garza:2015:22412} and
\ac{l-DFVB},\cite{Ying:2019:225}  where $\lambda$ refers to the hybrid
parameter that controls the admixture of nonlocal exchange effects.
Through this parameter, both \ac{pCCD-lDFT} and \ac{l-DFVB} can
interpolate between DFT and a \ac{MR} scheme (\ie, pCCD or VB).
\cite{Toulouse:2011:101102,Sharkas:2011:064113} In this work, we follow
the general strategy proposed in Refs.~\citenum{Sharkas:2012:044104}
and~\citenum{Kalai:2018:164105} within the context of the v2RDM-driven
formulation of MCPDFT, and we term the resulting approach \acs{lMCPDFT}.
\acs{lMCPDFT} differs from the \ac{MC1H} approach that inspired it in two
respects.  First, \acs{lMCPDFT} is not a self-consistent theory.  Second,
like \ac{pCCD-lDFT} (and unlike \ac{MC1H}), \acs{lMCPDFT}  relies on
XC functionals of the total and on-top pair densities, rather than the total 
and spin densities.




The manuscript is organized as follows. Section \ref{SEC:THEORY} provides
the theoretical details of \ac{lMCPDFT}. In Sec.~\ref{SEC:RESULTS}, we
discuss the application of \ac{lMCPDFT} to the dissociation of molecular
nitrogen and the symmetric double dissociation of a water molecule, as
well as to the standard O3ADD6 benchmark set. Concluding remarks and
potential future directions are then provided in
Sec.~\ref{SEC:CONCLUSION}.
	
	\section{Theory}
\label{SEC:THEORY}

Throughout this work, we use the conventional notation of \ac{MR} methods
when labeling the orbitals: the indices $i$, $j$, $k$, and $l$ denote
inactive (doubly occupied) orbitals; $t$, $u$, $v$, and $w$ represent
active orbitals; and $p$, $q$, $r$, and $s$ indicate general orbitals. A
summation over repeated indices is implied in all expressions.



We begin by defining the non-relativistic \acl{BO} electronic Hamiltonian
\begin{equation}
\hat{H} = h^p_q \hat{a}^\dag_{p_\sig} \hat{a}_{q_\sig} + \frac{1}{2} \nu^{pq}_{rs} \hat{a}^\dag_{p_\sig} \hat{a}^\dag_{q_\tau} \hat{a}_{s_\tau} \hat{a}_{r_\sig}
\end{equation}
where $\hat{a}^\dag$ and $\hat{a}$ represent second-quantized creation and
annihilation operators, respectively, and the Greek labels run over $\al$
and $\be$ spins (the sum over which is implied). The symbol 
$h^p_q = \braket{\psi_p|\hat{h}|\psi_q} $ represents the sum of the electron
kinetic energy and electron-nucleus potential energy integrals, and
$\nu^{pq}_{rs} = \braket{\psi_p \psi_q|\psi_r\psi_s}$ is an element of the
two-electron repulsion integral tensor. Because the electronic Hamiltonian
includes up to only pair-wise interactions, the ground-state energy of a
many-electron system can be expressed as an exact linear functional of the
the one-electron \ac{RDM} (\acs{1-RDM}) and two-electron \ac{RDM} (\acs{2-RDM})
\cite{Husimi:1940:264,Mayer:1955:1579,Lowdin:1955:1474}
\begin{equation}\label{EQ:Eel}
E = {}^1D^p_q h^p_q + \frac{1}{2} {}^2D^{pq}_{rs} \nu^{pq}_{rs}.
\end{equation}
\acused{1-RDM}\acused{2-RDM}Here, the \ac{1-RDM} and the \ac{2-RDM} are represented in their spin-free forms, 
with elements defined as
\begin{equation}
{}^1D^p_q = {}^1D^{p_\sigma}_{q_\sigma} = \braket{\Psi|\hat{a}^\dag_{p_\sig} \hat{a}_{q_\sig}|\Psi}	\label{EQ:1RDM}		\\
\end{equation}
and
\begin{equation}
{}^2D^{pq}_{rs} = {}^2D^{p_\sigma q_\tau}_{r_\sigma s_\tau} = \braket{\Psi|\hat{a}^\dag_{p_\sig} \hat{a}^\dag_{q_\tau} \hat{a}_{s_\tau} \hat{a}_{r_\sig}|\Psi} \label{EQ:2RDM},
\end{equation}
respectively. Again, the summation over the spin labels 
in Eqs. \ref{EQ:1RDM} and \ref{EQ:2RDM} is implied.

The \ac{MCPDFT} expression for the electronic energy is
\small
\begin{equation}\label{EQ:EMCPDFT}
\begin{aligned}
\EMCPDFT &= 2h^i_i + h^t_u {}^1D^t_u + E_\text{H} \\ 
         &+ E_\text{xc}\left[\rho,\Pi,|\nabla\rho|,|\nabla\Pi|\right],
\end{aligned}
\end{equation}
\normalsize
where $E_\text{xc}$ is the translated (t)\cite{LiManni:2014:3669} or
fully-translated (ft)\cite{Carlson:2015:4077} \ac{OTPD} \ac{XC} functional,
and the Hartree energy, $E_\text{H}$, is
\begin{equation}
E_\text{H} = 2 \nu^{ij}_{ij} + 2\nu^{ti}_{ui} {}^1D^t_u + \frac{1}{2} \nu^{tv}_{uw} {}^1D^{t}_{u} {}^1D^{v}_{w}
\end{equation}
The total electronic density and its gradient
that enter $E_\text{xc}$
are defined as
\begin{equation}
\label{EQ:RHO}  
\rho(\br) = {}^1D^p_q\ \psi^*_p(\br) \psi_q(\br),
\end{equation}
and
\begin{equation}
\label{EQ:DRHO}
\nabla\rho(\br) = {}^1D^p_q \left[ \nabla\psi^*_p(\br) \psi_q(\br) + \psi^*_p(\br) \nabla\psi_q(\br) \right],
\end{equation}
respectively.
The \ac{OTPD} and its gradient can similarly be expressed in terms of the \ac{2-RDM} as
\begin{equation}
\label{EQ:PI}
\Pi(\br)  = {}^2D^{pq}_{rs}\ \psi^*_p(\br) \psi^*_q(\br) \psi_r(\br) \psi_s(\br),
\end{equation}
and
\begin{eqnarray}
\label{EQ:DPI}  
\nabla\Pi(\br) = {}^2D^{pq}_{rs} &[& \nabla\psi^*_p(\br) \psi^*_q(\br) \psi_r(\br) \psi_s(\br) \nonumber \\
&+& \psi^*_p(\br) \nabla\psi^*_q(\br) \psi_r(\br) \psi_s(\br) \nonumber \\
&+& \psi^*_p(\br) \psi^*_q(\br) \nabla\psi_r(\br) \psi_s(\br) \nonumber \\
&+& \psi^*_p(\br) \psi^*_q(\br) \psi_r(\br) \nabla\psi_s(\br) ~],
\end{eqnarray}
respectively. In the \ac{MCPDFT} formalism, the \ac{1-RDM} and \ac{2-RDM}
can be taken from any reference calculation capable 
of generating ``good'' RDMs that include nondynamical
correlation effects.  In this work, the \ac{1-RDM} and \ac{2-RDM}
are taken from v2RDM-driven CASSCF calculations, and they
satisfy either two-particle (PQG)\cite{Garrod:1964:1756} or
PQG plus partial three-particle (T2)\cite{Zhao:2004:2095,Erdahl:1978:697}
$N$-representability conditions.

In order to reduce delocalization error in the \ac{MCPDFT} formalism, 
we define the \ac{lMCPDFT} energy expression
in which a fraction $\la \in (0,1)$ of the \ac{OTPD} exchange is replaced
with contributions from the reference 1- and \ac{2-RDM}:
\small
\begin{equation}\label{EQ:ElMCPDFT}
\begin{aligned}
\ELMCPDFT &= 2 h^i_i + h^t_u {}^1D^t_u + E_\text{H}	 \\
&+ \la (\frac{1}{2} \nu^{tv}_{uw} [{}^2\Delta^{tv}_{uw} - {}^1D^t_w {}^1D^v_u] - \nu^{tu}_{ii} {}^1D^t_u  - \nu^{ii}_{jj}) \\
& + \bar{E}^\la_\text{xc}\left[\rho,\Pi,|\nabla\rho|,|\nabla\Pi|\right],
\end{aligned}
\end{equation}
\normalsize
Here, the two-cumulant,
${}^2\Delta$,
is the part of the \ac{2-RDM} that cannot be represented 
in terms of the 1-RDM and is defined according to
\begin{equation}
{}^2D^{tv}_{uw} = {}^2\Delta^{tv}_{uw} + {}^1D^t_u {}^1D^v_w - {}^1D^t_w {}^1D^v_u.
\end{equation}
The complement \ac{OTPD} functional, $\bar{E}^\la_\text{xc}[\cdot]$,
is given by
\small
\begin{equation}
\begin{aligned}
\bar{E}^\la_\text{xc}[\rho,\Pi,|\nabla\rho|,|\nabla\Pi|]  &=	E_\text{xc}[\rho,\Pi,|\nabla\rho|,|\nabla\Pi|] 	\\
&- E^\la_\text{xc}[\rho,\Pi,|\nabla\rho|,|\nabla\Pi|]
\end{aligned}
\end{equation}
\normalsize
which is the difference between the conventional \acl{XC} \ac{OTPD}
functional ($E_\text{xc}[\cdot]$) and its $\la$-dependent hybrid
version ($E^\la_\text{xc}[\cdot]$).\cite{Sharkas:2012:044104}  

The exchange part of the complement functional is easily defined, 
as it simply scales linearly with the mixing parameter $\la$
\cite{Kalai:2018:164105,Sharkas:2011:064113,Sharkas:2012:044104}
\small
\begin{equation}\label{EQ:EHX}
	\bar{E}^\la_\text{x}[\rho,\Pi,|\nabla\rho|,|\nabla\Pi|] = (1-\la) E_\text{x}[\rho,\Pi,|\nabla\rho|,|\nabla\Pi|]
\end{equation}
\normalsize
The correlation contribution to the complement \ac{OTPD} functional can be obtained through uniform coordinate scaling of the density as \cite{Kalai:2018:164105,Sharkas:2011:064113,Sharkas:2012:044104}
\footnotesize
\begin{equation}\label{EQ:EC1}
	\begin{aligned}
		&\bar{E}^\la_\text{c}[\rho,\Pi,|\nabla\rho|,|\nabla\Pi|]  =	E_\text{c}[\rho,\Pi,|\nabla\rho|,|\nabla\Pi|] - E^\la_\text{c}[\rho,\Pi,|\nabla\rho|,|\nabla\Pi|]		\\
		&= E_\text{c}[\rho,\Pi,|\nabla\rho|,|\nabla\Pi|] - \la^2 E_\text{c}[\rho^{1/\la},\Pi^{1/\la},|\nabla\rho^{1/\la}|,|\nabla\Pi^{1/\la}|]
	\end{aligned}
\end{equation}
\normalsize
in which the scaled density and \ac{OTPD} functions are defined as \cite{Higuchi:2013:91}
\begin{gather}
	\rho^{1/\la}(\br) = (1/\la^3)\ \rho(\br/\la) 	\label{EQ:SCALEDRHO}	\\	
	\Pi^{1/\la}(\br) = (1/\la^6)\ \Pi(\br/\la)	\label{EQ:SCALEDPI}
\end{gather}
Following Ref.~\cite{Sharkas:2012:044104}, we neglect the scaling
relations of the total density and \ac{OTPD} in the correlation complement
functional and define
\begin{equation}\label{EQ:EC2}
	\bar{E}^\la_\text{c}[\rho,\Pi,|\nabla\rho|,|\nabla\Pi|]  =	(1 - \la^2) E_\text{c}[\rho,\Pi,|\nabla\rho|,|\nabla\Pi|]
\end{equation}
Using Eqs.~\ref{EQ:EHX} and \ref{EQ:EC2}, the \ac{lMCPDFT} energy
expression takes its final form:
\begin{equation}\label{EQ:ElMCPDFT2}
\begin{aligned}
\ELMCPDFT &= 2h^i_i + h^t_u {}^1D^t_u + E_\text{H}		\\
		  &+ \la (\frac{1}{2}\nu^{tv}_{uw} [{}^2\Delta^{tv}_{uw} - {}^1D^t_w {}^1D^v_u] - \nu^{tu}_{ii} {}^1D^t_u - \nu^{ii}_{jj}) 	\\ 
		  &+ (1-\la) E_\text{x}[\rho,\Pi,|\nabla\rho|,|\nabla\Pi|]		\\
		  &+ (1-\la^2) E_\text{c}[\rho,\Pi,|\nabla\rho|,|\nabla\Pi|]
\end{aligned}
\end{equation}
where, for $\la=0$ or $\la=1$, the \ac{lMCPDFT} energy expression reduces
to that corresponding to \ac{MCPDFT} or the underlying multiconfiguration
reference method, respectively.

	\section{Results and Discussion}
	
\label{SEC:RESULTS}

The RDMs that enter the MCPDFT and \ac{lMCPDFT} equations were obtained
from v2RDM-driven CASSCF calculations performed using a plugin
\cite{V2RDM:GITHUB} to the \textsc{Psi4} electronic structure
package;\cite{Turney:2012:1759} the optimized RDMs satisfy either the PQG
or PQG+T2 $N$-representability conditions.  Both the \ac{MCPDFT} and
\ac{lMCPDFT} procedures, along with the translated and
fully-translated versions of the \acs{SVWN3},
\cite{Gaspar:1974:213,Slater:1951:385,Vosko:1980:1200} \acs{PBE}
\cite{Perdew:1996:3865}, and \acs{BLYP}
\cite{Becke:1988:3098,Lee:1988:785} \ac{XC} functionals, have also been
implemented in our open-source library
(\textsc{OpenRDM})\cite{OPENRDM:GITHUB} which is interfaced as a plugin to
\textsc{Psi4}.\cite{RDMINOLES:GITHUB}


\subsection{\ce{N2} and \ce{H2O} Bond Dissociations}\label{SUBSEC:PBES}

In this section, we consider the accuracy of \ac{MCPDFT} and \ac{lMCPDFT}
for describing the dissociation of \ce{N2} and the double dissociation of
\ce{H2O}; the \acp{NPE} in the respective \acp{PEC}, which are obtained by
comparing the \ac{MCPDFT} and \ac{lMCPDFT} results with those from
\ac{CASPT2}, serve as a useful metric in this context.  The \ac{CASSCF}
calculations underlying \ac{CASPT2}, \ac{MCPDFT}, and \ac{lMCPDFT}
employed full-valence active spaces and the cc-pVTZ basis
set,\cite{Dunning:1989:1007} and, in the case of v2RDM-driven CASSCF, the
electron repulsion integral (ERI) tensor was represented within the \ac{DF}
approximation,\cite{Whitten:1973:4496,Dunlap:1979:3396} using the
cc-pVTZ-JK auxiliary basis set.\cite{Weigend:2002:4285} 
All \ac{MCPDFT} and \ac{lMCPDFT} computations also employed the DF
approximation.  The \ac{CASPT2} computations were performed using the
\textsc{Open-MOLCAS} electronic structure
package,\cite{Aquilante:2016:506} and the standard imaginary shift
\cite{Forsberg:1997:196} of 0.20 E$_{\rm h}$ and \textsc{Open-MOLCAS}'s
default value of 0.25 E$_{\rm h}$ for the IPEA shift\cite{Ghigo:2004:142}
were applied throughout.

\begin{figure}[!htbp]
	\setlength{\abovecaptionskip}{-1pt}	

	\caption{The \acs{NPE} in the dissociation curves for the \ce{N2}
	molecule as a function of the mixing parameter, $\la$, when using
	PQG (a),(b) or PQG+T2 (c),(d) $N$-representability conditions and
	translated (a),(c) or fully-translated (b),(d) \acs{lMCPDFT}
	functionals.}

	\centering
	\includegraphics{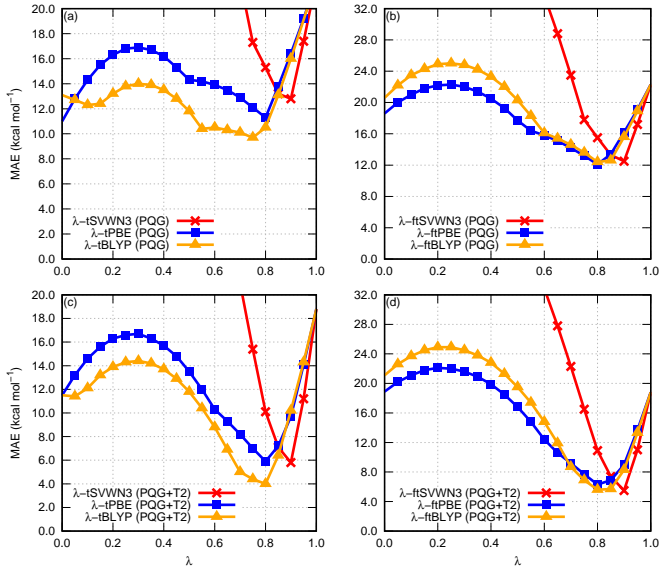}
	\label{FIG:N2}
\end{figure}

Figure \ref{FIG:N2} illustrates the effect of the mixing parameter, $\la$,
on the \acs{NPE} for \ce{N2} when enforcing the  PQG or PQG+T2
$N$-representability conditions and adopting either translated or
fully-translated \acs{lMCPDFT} functionals. The \ac{NPE} is defined as the
difference in the maximum and the minimum deviations between the
\ac{lMCPDFT} and \ac{CASPT2} \acp{PEC} from an N--N distance of 0.7 \AA\
to an N--N distance of 5.0 \AA.  With the exception of that for tBLYP
combined with the PQG $N$-representability conditions, all \ac{NPE} curves
presented in Fig. \ref{FIG:N2} exhibit their minimal value between
$\la=0.70$ and $\la=0.90$.  Specifically, when enforcing the PQG (PQG+T2)
$N$-representability conditions and adopting \acs{tSVWN3}, \acs{tPBE},
\acs{tBLYP}, \acs{ftSVWN3}, \acs{ftPBE}, and \acs{ftBLYP} functionals, the
optimal $\la$ values are 0.90 (0.90), 0.00 (0.80), 0.75 (0.80), 0.90 (0.90),
0.80 (0.80), and 0.80 (0.80), respectively. For tPBE/PQG, it appears that the
optimal $\la$ value is 0.00 in this case, although the NPE at $\la$ = 0.80 is only
0.3 kcal mol$^{-1}$ larger.  Note that the optimal values of the mixing
parameters in all other cases lie between $\la=0.70$ and $\la=0.90$, which
are close to the value of $\la=0.75$ used within the \ac{pCCD-lDFT}-based study of \ce{N2}
dissociation presented in Ref. \citenum{Garza:2015:22412}.


\begin{figure}[!htbp]
	\setlength{\abovecaptionskip}{-2pt}

	\caption{\ac{MCPDFT} and \ac{lMCPDFT} \acp{NPE} (mhartree) associated with
	the dissocation of \ce{N2}.  The mixing parameter for \ac{lMCPDFT}
	is chosen to be that which gives the lowest \ac{NPE} for each
	functional considered in Fig. \ref{FIG:N2}. Each concentric ring denotes an additional 15 mhartree error.}

	\centering
	\includegraphics{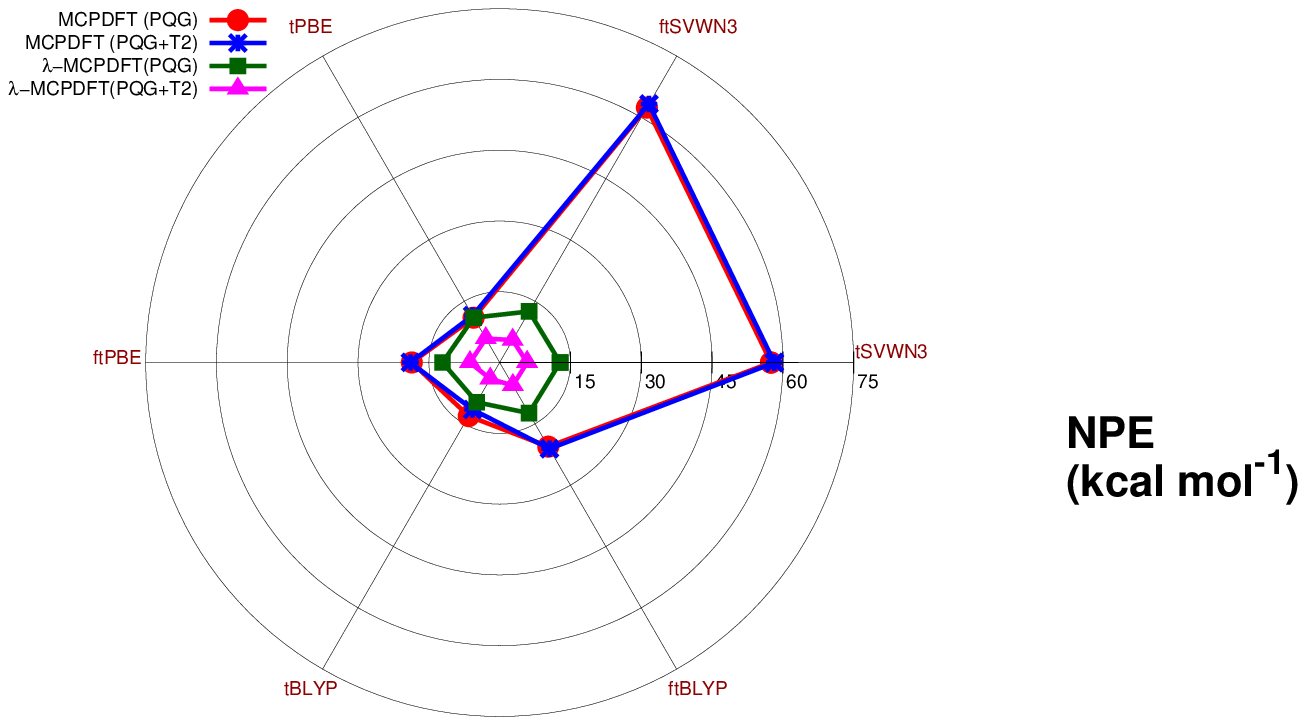}
	\label{FIG:N2POLAR}
\end{figure}

Figure \ref{FIG:N2POLAR} presents the \acp{NPE} associated with the
\ac{MCPDFT} and hybrid \ac{lMCPDFT} \acp{PEC}, using the optimal $\la$
values identified for each functional / $N$-representability combination
considered in Fig.  \ref{FIG:N2}, and the data reveal several interesting
features.  First, as noted in Ref. \citenum{Mostafanejad:2019:290}, for
both translated and fully-translated \ac{MCPDFT} functionals, the \ac{NPE}
is somewhat insensitive to the $N$-representability of the RDMs.  On the
other hand, \ac{lMCPDFT} clearly yields lower \ac{NPE} values when the
underlying RDMs satisfy both the PQG and T2 conditions; it appears that
the consideration of nonlocal exchange effects alleviates enough of the
delocalization error to allow us to discern the error associated with
approximate $N$-representability of the reference RDMs. Second, while the
\acp{NPE} vary significantly with the choice of \ac{MCPDFT} functional,
nonlocal exchange effects serve as a great equalizer; the \ac{NPE} are
quite similar for all \ac{lMCPDFT} methods, when enforcing a given set of
$N$-representability conditions.  Unsurprisingly, the improvement in the
\acp{NPE} is largest for the \acused{ltSVWN3}\ac{ltSVWN3} and
\acused{lftSVWN3}\ac{lftSVWN3} functionals.  Third, as observed in Ref.
\citenum{Mostafanejad:2019:290}, the choice of translated or
fully-translated functionals does not significantly affect the \ac{NPE}
associated with functionals within either the \ac{MCPDFT} or \ac{lMCPDFT}
formalisms.  Fully-translated functionals yield slightly lower \ac{NPE}
values than translated functionals, but this difference is far less
significant than the improvements afforded by the admixture of nonlocal
exchange effects.

\begin{figure}[!htbp]
	\setlength{\abovecaptionskip}{-2pt}

	\caption{\ac{MCPDFT} and \ac{lMCPDFT} \acp{NPE} (mhartree) associated with
	the double dissociation of \ce{H2O}. The mixing parameter for
	\ac{lMCPDFT} is chosen to be that which gives the lowest \ac{NPE}
	for each functional (see Fig. S1 of the Supporting Information).  Each concentric ring denotes an additional 15 mhartree error.}

	\centering
	\includegraphics{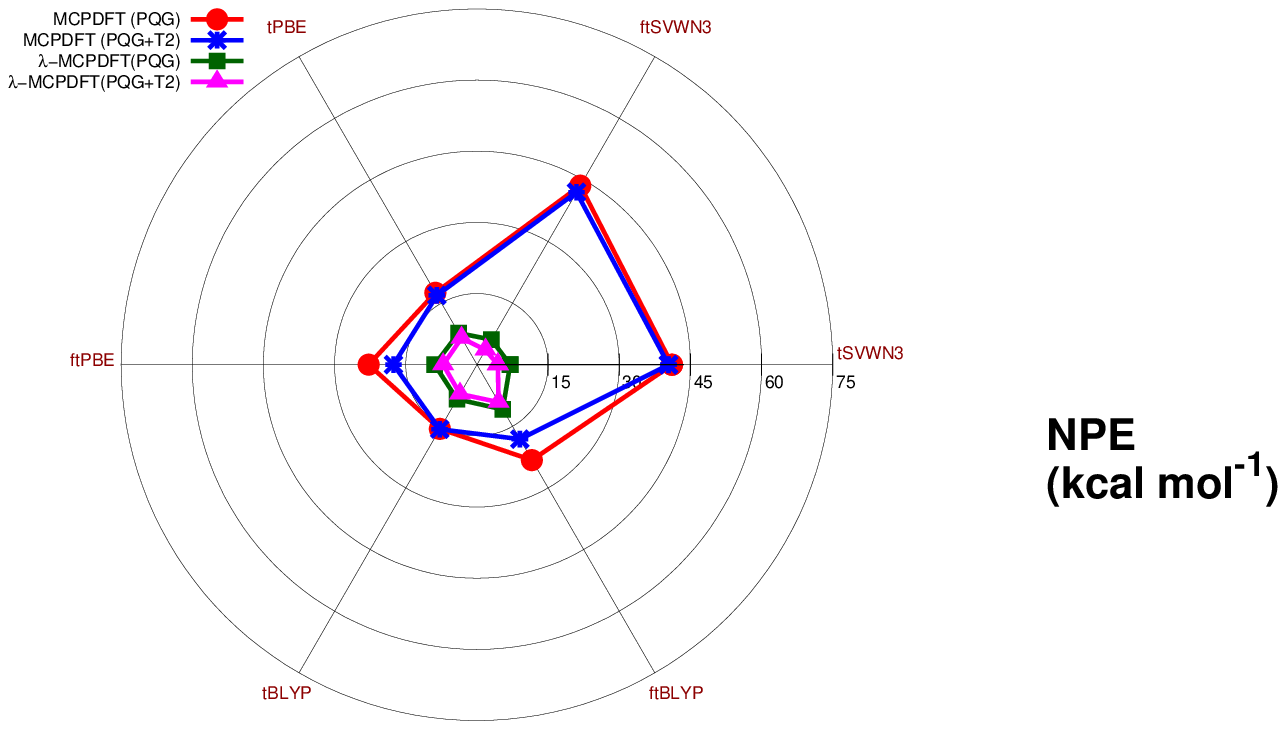}
	\label{FIG:H2OPOLAR}
\end{figure}

Similar trends are observed for the \acp{NPE} associated with the \ce{H2O}
double-dissociation \ac{PEC}.  Figure \ref{FIG:H2OPOLAR} depicts the
\ac{NPE} for each functional, where the mixing parameter is chosen to
minimize the \ac{lMCPDFT} \ac{NPE} between O--H bond distances of 0.6 \AA\
and 5.0 \AA.  In this case, when enforcing the PQG (PQG+T2)
$N$-representability conditions and adopting \acs{tSVWN3}, \acs{tPBE},
\acs{tBLYP}, \acs{ftSVWN3}, \acs{ftPBE}, and \acs{ftBLYP} functionals, the
optimal $\la$ values are 0.70 (0.70), 0.45 (0.50), 0.45 (0.50), 0.70 (0.70),
0.50 (0.50), and 0.50 (0.50), respectively.  Relative to the case of \ce{N2}, the
optimal mixing parameters, in general, are smaller for this problem.  The
only exception is tPBE, where the optimal value of $\la=0.45$ is clearly
larger than the optimal value for \ce{N2} dissociation (0.0). We again
note that the inclusion of nonlocal exchange effects significantly reduces
the \acp{NPE}, in general, and the variation in the performance of
\ac{lMCPDFT} functionals is much less than that of the \ac{MCPDFT}
functionals.  In constrast to the case of \ce{N2}, the quality of the
\acp{NPE} associated with \ac{lMCPDFT} functionals is less dependent upon
the $N$-representability of the underying RDMs. The NPEs for each \ac{lMCPDFT}
functional, as a function of the mixing parameter, are depicted in Fig.
S1 of the Supporting Information.

\subsection{O3ADD6 Benchmark Set}\label{SUBSEC:O3ADD6}


The O3ADD6 benchmark set is comprised of the energies of three stationary
points associated with the 1,3-dipolar cycloadditions of ozone (\ce{O3}) to
ethylene (\ce{C2H4}) and \ce{O3} to acetylene (\ce{C2H2}), relative to the
energies of the isolated reactants.  The stationary points [the van der
Waals complex (vdW), the \ac{TS}, and the cycloadduct (Cycloadd.)] and the
separated reactants are all assumed to be in singlet spin states. These
systems present a challenge to quantum chemical methods, as the \ac{HOMO}
$\to$ \ac{LUMO} double excitation character in \ce{O3} and the
near-degeneracy of the $\pi$ and $\pi^*$ orbitals corresponding to the
reactive $\pi$ bonds in the olefins result in strong \ac{MR} character in
several of the species along the reaction
coordinate.\cite{Wheeler:2008:1798,Sharkas:2012:044104}  Here, we apply
the \ac{MCPDFT} and \ac{lMCPDFT} methods to this data set, using the
\acs{aug-cc-pVTZ} basis set\cite{Dunning:1989:1007} and geometries for the
stationary points given in Ref.~\citenum{Zhao:2009:5786}.  All
v2RDM-driven CASSCF and \ac{lMCPDFT} calculations employed the DF
approximation to the ERI tensor and the aug-cc-pVTZ-JK auxiliary basis
set. In order to compare our results with those from
Refs.~\citenum{Sharkas:2012:044104} and \citenum{Zhao:2009:5786}, we
consider an active space comprised of 2 electrons in 2 orbitals [denoted
as (2e,2o)] for the reactants and (4e,4o) active spaces  for the vdW,
\ac{TS}, and Cycloadd.~species; the orbitals comprising each active space
are defined in Ref.  \citenum{Sharkas:2012:044104}.  Following
Refs.~\citenum{Zhao:2009:5786} and \citenum{Sharkas:2012:044104}, we
neglect vibrational zero-point energy contributions to the energies of
the stationary points and separated reactants.

%
%

\begin{figure}[!ht]

\caption{The \acs{MAE} (kcal mol$^{-1}$) in calculated O3ADD6 energies as
a function of the mixing parameter, $\la$, when RDMs satisfy PQG (a),(b)
and PQG+T2 (c),(d) $N$-representability conditions and when adopting
translated (a),(c) and fully-translated (b),(d) \acs{lMCPDFT}
functionals.}

	\label{FIG:MAEVSLAMBDA}
	\centering
	\includegraphics{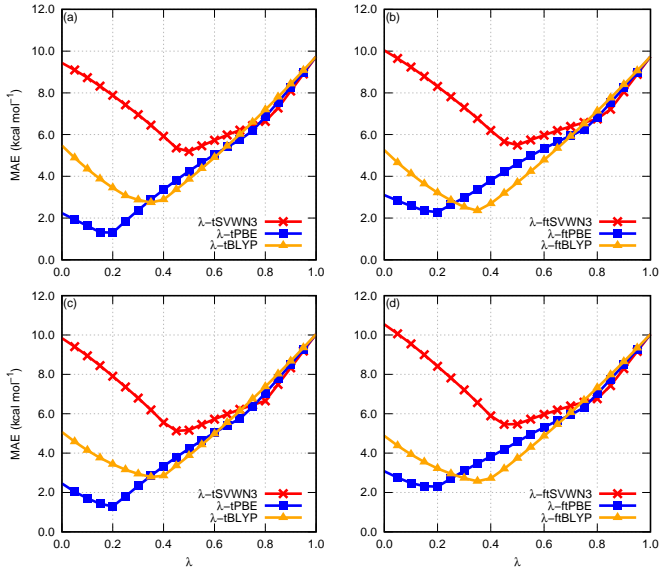}
\end{figure}

We identify the optimal nonlocal exchange mixing parameter, specific to
the O3ADD6 benchmark set, for several \ac{lMCPDFT} functionals by
minimizing the \ac{MAE} in the relative energies of the stationary points
and the isolated reactant molecules. The reference values to which the
calculated relative energies are compared are taken from
Ref.~\citenum{Zhao:2009:5786}.  Fig.~\ref{FIG:MAEVSLAMBDA} depicts the
\acp{MAE} in the calculated relative energies as a function of $\la$, and
the data clearly convey the importance of nonlocal exchange effects for
this problem, particularly in the case of $\la$-tSVWN3 and $\la$-ftSVWN3.
For all functionals considered, the inclusion of some fraction of nonlocal
exchange is beneficial, although this fraction is, in general, smaller
than the optimal fraction for minimizing the \ac{NPE} in the \ce{N2} and
\ce{H2O} dissociation curves considered above. The optimal $\la$ values,
which are tabulated in Table \ref{TAB:O3ADD6} and in Table S1 in the
Supporting Information, are insensitive to the choice of
$N$-represetability conditions.  The only functional that requires a
different mixing parameter under different $N$-representability conditions
is \acs{tSVWN3}, and this value only changes by 0.05 in this case.  One
possible reason for this insensitivity is the small size of the active
spaces considered.  For example, for the (2e,2o) active space, the
two-particle conditions alone are sufficient to yield ensemble-state
$N$-representable RDMs.


\begin{table*}[!htbp]
	\centering
	\setlength{\tabcolsep}{4pt}
	\setlength{\extrarowheight}{1pt}

	\caption{Calculated relative energies (kcal mol$^{-1}$) of the
	stationary points and separated reactant molecules that comprise
	the O3ADD6 dataset.}

	\label{TAB:O3ADD6}
	\resizebox{\textwidth}{!}{
		\begin{tabular}{lcccccccccc}
			\hline\hline
			\mr{2}{*}{Method}    & \mr{2}{*}{$N$-representability} & \mr{2}{*}{$\la$$^a$} & \mc{3}{c}{\ce{O3 + C2H2 $\rightarrow$}} & \mc{3}{c}{\ce{O3 + C2H4 $\rightarrow$}} & \mr{2}{*}{\acs{MAE}} &        \\
			\cline{4-6}\cline{7-9}
			&                                 &                  &                   vdW                   &                   TS                    &     Cycloadduct      &  vdW  &  TS   & Cycloadduct &       &  \\
			\hline
			$\la$-tPBE                 &         \mr{5}{*}{PQG}          &       0.20       &                  -0.42                  &                  7.84                   &        -67.75        & -2.19 & 4.91  &   -57.66    & 1.30  &  \\
			$\la$-tBLYP                &                                 &       0.35       &                  0.24                   &                  12.51                  &        -61.41        & -2.70 & 8.82  &   -56.18    & 2.75  &  \\
			$\la$-tSVWN3               &                                 &       0.50       &                  -1.15                  &                  7.72                   &        -73.88        & -5.72 & 4.69  &   -72.34    & 5.19  &  \\
			v2RDM-CASSCF         &                                 &       1.00       &                  0.55                   &                  23.33                  &        -56.41        & -6.93 & 18.97 &   -69.54    & 9.73  &  \\[3pt]
			$\la$-tPBE                 &        \mr{5}{*}{PQG+T2}        &       0.20       &                  -0.40                  &                  7.69                   &        -68.00        & -1.86 & 4.87  &   -57.57    & 1.29  &  \\
			$\la$-tBLYP                &                                 &       0.35       &                  0.27                   &                  12.82                  &        -61.51        & -1.97 & 9.23  &   -55.88    & 2.78  &  \\
			$\la$-tSVWN3               &                                 &       0.45       &                  -1.26                  &                  6.92                   &        -75.19        & -4.51 & 4.10  &   -71.78    & 5.13  &  \\
			v2RDM-CASSCF         &                                 &       1.00       &                  0.63                   &                  25.86                  &        -55.60        & -4.19 & 21.55 &   -68.13    & 10.04 &  \\[3pt]
			MC1H-PBE$^b$         &         \mr{2}{*}{---}          &       0.25       &                  -1.08                  &                  3.66                   &        -70.97        & -1.25 & 0.13  &   -61.26    & 3.35  &  \\
			MC1H-BLYP$^b$        &                                 &       0.25       &                  -0.36                  &                  6.74                   &        -63.76        & -0.47 & 2.57  &   -54.21    & 1.30  &  \\
			\hline
			Reference values$^c$ &                                 &       ---        &                  -1.90                  &                  7.74                   &        -63.80        & -1.94 & 3.37  &   -57.15    &  ---  &  \\
			\hline\hline
		\end{tabular}
	}
	\begin{tablenotes}
		\scriptsize
		\item $^a$ The optimal mixing parameter.
		\item $^b$ From Ref.~\citenum{Sharkas:2012:044104}.
		\item $^c$ Best estimates from Ref.~\citenum{Zhao:2009:5786}.
	\end{tablenotes}
\end{table*}

While the v2RDM-driven CASSCF method can capture nondynamical correlation
effects that are important in the systems comprising the O3ADD6 set, the
large average errors associated with the approach (9.73 and 10.04 kcal
mol$^{-1}$ when the calculations are performed under the PQG and PQG+T2
$N$-representability conditions, respectively) reflect its inability to
describe dynamical electron correlation. These effects are well described
by \ac{MCPDFT} and \ac{lMCPDFT}; each hybrid functional considered here,
as well as the base \ac{MCPDFT} functionals considered in the Supporting
Information (with the exception of the translated and fully-translated
SVWN3 functionals), substantially reduce the error associated with CASSCF.
In particular, the admixture of 20\% nonlocal exchange into the translated
\acs{PBE} \ac{OTPD} functional reduces the \ac{MAE} to only 1.30 kcal
mol$^{-1}$ (PQG) or 1.29 kcal mol$^{-1}$ (PQG+T2), which is close to the
1.0 kcal mol$^{-1}$ threshold for ``chemical accuracy.'' The description
of the O3ADD6 afforded by \ac{lMCPDFT} with this functional  is similar in
quality to that exhibited by the \ac{MC1H}-\acs{BLYP} approach of
Ref.~\citenum{Sharkas:2012:044104} and superior to that displayed by the
\ac{MC1H}-\acs{PBE} approach of that same work.

We note that the quality of the \ac{lMCPDFT} results, like that of the
v2RDM-driven CASSCF, is independent of the $N$-representabilty of the
underlying RDMs, in this case.  As mentioned above, the likely reason for
this behavior is the small size of the active spaces employed.  Both sets
of conditions we consider yield exact CASSCF RDMs for systems with (2e,2o)
active spaces, and the (4e,4o) active spaces are not so large that we
would expect significantly differences to arise between RDMs satisfying
the PQG  and PQG+T2 conditions.  We also note that full translation of the
OTPD functionals does not necessarily lead to a reduction in the MAEs
associated with 
\ac{MCPDFT} or \ac{lMCPDFT}.  For example, a
fully-translated PBE functional with a 20\% admixture of nonlocal exchange
(the optimal value for this fully-translated functional) performs slightly
worse than the corresponding translated functional; the MAE in this case
is 2.28 kcal mol$^{-1}$.  A complete set of O3ADD6 relative energies and
MAEs for translated and fully-translated \ac{MCPDFT} and \ac{lMCPDFT} can
be found in the Supporting Information.

	\section{Conclusions}

\label{SEC:CONCLUSION}

We have presented a one-parameter global hybrid extension of the
multiconfigurational pair density functional theory approach, inspired by
the multiconfigurational one-parameter hybrid DFT developed in
Ref.~\citenum{Sharkas:2012:044104}.  Like MC1H, \ac{lMCPDFT} relies upon a
linear decomposition of the electron-electron interaction operator to
incorporate a fraction of nonlocal exchange into the MCPDFT formalism.
Unlike MC1H, \ac{lMCPDFT} is not a self-consistent theory, and the
functionals we employ are functionals of the on-top pair density, whereas
the those employed in Ref.~\citenum{Sharkas:2012:044104} are standard
functionals of the spin density. In both of these respects, the present
approach is similar to the \ac{pCCD-lDFT} of
Ref.~\citenum{Garza:2015:044109}, with the principal distinction unique to
this work being the source of the reference on-top pair densities. Not
surprisingly, the admixture of nonlocal exchange improves the quality of
the \ac{MCPDFT} energy, as measured by NPEs in molecular dissociation
curves and the performance of the approach when applied to the O3ADD6
benchmark data set.

The results presented herein represent a best-possible scenario for the
accuracy of \ac{lMCPDFT}, in that the mixing parameter is system dependent
and chosen specifically to minimize the error for the case in question.  A
universally useful \ac{lMCPDFT} formalism would require a single mixing
parameter optimized over a large data set.  Alternatively, the accuracy of
the approach could be improved, in principle, by choosing a local mixing
parameter,\cite{MIEHLICH1997,Grafenstein2005,Grafenstein2000,Gritsenko:2018:062510,Gritsenko:2019:024111}
$f(\br)$, as opposed to a constant value $\la$, which could be defined in
terms of the total density and the \ac{OTPD}. In addition, the MCPDFT
formalism could be futher improved by generalizing the OTPD functionals to
consider range separation of the coulomb interaction (see
Ref.~\citenum{Garza:2015:22412}, for example) or double hybrization.

Lastly, we highlight one unfortunate formal aspect of \ac{lMCPDFT}.  One
of the nice properties of MCPDFT is that, unlike some other MR+DFT
methods, it avoids double counting of electron correlation within the
active space.  However, upon introducing a non-zero exchange mixing
parameter, $\la$, this property is formally lost in \ac{lMCPDFT}.  The
two-cumulant in Eq. \ref{EQ:ElMCPDFT} introduces some correlation effects
within the active space that may also be described by the OTPD functional.
Presumably, this double counting is minimized along with the total error
in any fitting procedure used to determine $\la$, but we cannot formally
guarantee that this is the case.

	\vspace{0.5cm}
	{\bf Acknowledgments}
    This material is based upon work supported by the Army Research Office
    Small Business Technology Transfer (STTR) program under Grants No.
    W911NF-16-C-0124 and W911NF-19-C0048. Mohammad Mostafanejad was supported
    by a fellowship from The Molecular Sciences Software Institute under NSF
    grant ACI-1547580.
	
\label{SEC:ACRONYMS}

\begin{acronym}
	\acro{ZPE}{zero-point energy}
	\acro{TS}{transition state}	
	\acro{QTAIM}{quantum theory of atoms in molecules}
	\acro{DC-KDFT}{density corrected-\acl{KDFT}}
	\acro{KDFT}{kinetic \acl{DFT}}
	\acro{DC-DFT}{density corrected-\acl{DFT}}
	\acro{SI}{Supporting Information}
	\acro{RASSCF}{restricted active-space \acl{SCF}}
	\acro{GASSCF}{generalized active-space \acl{SCF}}
	\acro{CASSCF}{\acl{CAS} \acl{SCF}}
	\acro{ORMAS}{occupation-restricted multiple active-space}
	\acro{DE}{delocalization error}
	\acro{LE}{localization error}
	\acro{SIE}{self-interaction error}
	\acro{CC}{coupled-cluster}
	\acro{CS}{constrained search}
	\acro{CS-KSDFT}{\acl{CS}-\acl{KSDFT}}
	\acro{MO}{molecular orbital}
	\acro{BO}{Born-Oppenheimer}
	\acro{CPO}{correlated participating orbitals}
	\acro{SCF}{self-consistent field}
	\acro{CAS}{complete active-space}
	\acro{PT}{perturbation theory}
	\acro{PT2}{second-order \acl{PT}}
	\acro{ACI-DSRG-MRPT2}{\acl{ACI}-\acl{DSRG} \acl{MR} \acl{PT2}}
	\acro{CASPT2}{\acl{CAS} \acl{PT2}}
	\acro{IPEA}{ionization potential electron affinity}
	\acro{MC}{multiconfiguration}
	\acro{HF}{Hartree-Fock}
	\acro{MR}{multireference}
	\acro{MR-AQCC}{\acl{MR}-averaged quadratic \acl{CC}}
	\acro{WF}{wave function}
	\acro{CI}{configuration interaction}
	\acro{ACI}{adaptive \acl{CI}}
	\acro{FCI}{full \acl{CI}}
	\acro{ACSE}{anti-Hermitian contracted Schr\"odinger equation}
	\acro{DSRG}{driven similarity renormalization group}
	\acro{DMRG}{density matrix renormalization group}
	\acro{RPA}{random-phase approximation}
	\acro{pp-RPA}{particle-particle \acl{RPA}}
	\acro{CSE}{contracted Schr\"odinger equation}
	\acro{ACSE}{anti-Hermitian \acl{CSE}}
	\acro{aug-cc-pVTZ}{augmented correlation-consistent polarized-valence triple-$\ze$}
	\acro{aug-cc-pVQZ}{augmented correlation-consistent polarized-valence quadruple-$\ze$}
	\acro{aug-cc-pwCV5Z}[aug-cc-p$\om$CV5Z]{augmented correlation-consistent polarized weighted-core-valence quintuple-$\ze$}
	\acro{cc-pVTZ}{correlation-consistent polarized-valence triple-$\ze$}
	\acro{cc-pVDZ}{correlation-consistent polarized-valence double-$\ze$}
	\acro{WFT}{wave function theory}
	\acro{pCCD}{pair coupled-cluster doubles}
	\acro{pCCD-lDFT}[pCCD-$\la$DFT]{\acl{pCCD} $\la$\acs{DFT}}
	\acro{VB}{valence bond}
	\acro{l-DFVB}[$\la$-DFVB]{$\la$-density functional \acl{VB}}
	\acro{DF}{density-fitting}
	\acro{ERI}{electron-repulsion integral}
	\acro{ZPVE}{zero-point vibrational energy}
	\acro{MCSCF}{\acl{MC} \acl{SCF}}
	\acro{v2RDM-DOCI}{\acl{v2RDM}-\acl{DOCI}}
	\acro{PES}{potential energy surface}
	\acro{MCPDFT}{\acl{MC} \acl{PDFT}}
	\acro{MC1H}{\acl{MC} one-parameter hybrid \acl{DFT}}
	\acro{MCHPDFT}{\acl{MC} hybrid-\acl{PDFT}}
	\acro{lMCPDFT}[$\la$-MCPDFT]{\acl{MC} \acl{1HPDFT}}
	\acro{MCRSHPDFT}[$\mu\la$-MCPDFT]{\acl{MC} range-separated hybrid-\acl{PDFT}}	
	\acro{GMCPDFT}{generalized \acl{MCPDFT}}
	\acro{MC1HPDFT}{\acl{MC} \acl{1HPDFT}}
	\acro{ltPBE}[$\la$-\acs{tPBE}]{$\la$-\acl{tPBE}}
	\acro{ltrevPBE}[$\la$-\acs{trevPBE}]{$\la$-\acl{trevPBE}}
	\acro{ltBLYP}[$\la$-\acs{tBLYP}]{$\la$-\acl{tBLYP}}
	\acro{ltSVWN3}[$\la$-\acs{tSVWN3}]{$\la$-\acl{tSVWN3}}
	\acro{lftPBE}[$\la$-\acs{ftPBE}]{$\la$-\acl{ftPBE}}
	\acro{lftrevPBE}[$\la$-\acs{ftrevPBE}]{$\la$-\acl{ftrevPBE}}
	\acro{lftBLYP}[$\la$-\acs{ftBLYP}]{$\la$-\acl{ftBLYP}}
	\acro{lftSVWN3}[$\la$-\acs{ftSVWN3}]{$\la$-\acl{ftSVWN3}}
	\acro{LMF}{local mixing function}
	\acro{DOCI}{doubly occupied \acl{CI}}
	\acro{KSDFT}{\acl{KS} \acl{DFT}}
	\acro{DFT}{density functional theory}
	\acro{PDFT}{pair-\acl{DFT}}
	\acro{HPDFT}{hybrid \acl{PDFT}}
	\acro{1HPDFT}{one-parameter hybrid \acl{PDFT}}	
	\acro{OTPD}{on-top pair-density}
	\acro{ft}{full translation}
	\acro{tr}{conventional translation}
	\acro{RDM}{reduced-density matrix}
	\acrodefplural{RDM}{reduced-density matrices}
	\acro{1-RDM}{one-electron \acl{RDM}}
	\acro{2-RDM}{two-electron \acl{RDM}}
	\acrodefplural{2-RDM}{two-electron reduced-density matrices}
	\acro{3-RDM}{three-electron \acl{RDM}}
	\acrodefplural{3-RDM}{three-electron \aclp{RDM}}
	\acro{4-RDM}{four-electron \acl{RDM}}
	\acro{HRDM}{hole \acl{RDM}}
	\acro{FP-1}{frontier partition with one set of interspace excitations}
	\acro{SF}{spin-flip}
	\acro{CCSD}{coupled-cluster with singles and doubles}
	\acro{SF-CCSD}{\acl{SF}-\acl{CCSD}}
	\acro{1-HRDM}{one-hole \acl{RDM}}
	\acro{2-HRDM}{two-hole \acl{RDM}}
	\acro{PEC}{potential energy curve}
	\acro{CCSDT}{coupled-cluster, singles doubles and triples}
	\acro{KS}{Kohn-Sham}
	\acro{HK}{Hohenberg-Kohn}
	\acro{XC}{exchange-correlation}
	\acro{HXC}{Hartree-\acl{XC} }		
	\acro{LSDA}{local spin-density approximation}
	\acro{GGA}{generalized gradient approximation}
	\acro{NGA}{nonseparable gradient approximation}
	\acro{MP2}{second-order M\o ller-Plesset \acl{PT}}
	\acro{LO}{Lieb-Oxford}
	\acro{SVWN3}{Slater and Vosko-Wilk-Nusair random-phase approximation expression III}
	\acro{PBE}{Perdew-Burke-Ernzerhof}
    \acro{revPBE}{revised \acs{PBE}}
	\acro{PBE0}{hybrid-\acs{PBE}}
	\acro{MN15}{Minnesota 15}
	\acro{TPSS}{Tao-Perdew-Staroverov-Scuseria}
	\acro{LYP}{Lee-Yang-Parr}
	\acro{PW91}{Perdew-Wang 91}
	\acro{BLYP}{Becke and \acl{LYP}}
	\acro{B3LYP}{Becke-3-\acl{LYP}}
	\acro{tSVWN3}{translated \acl{SVWN3}}
	\acro{tPBE}{translated \acl{PBE}}
    \acro{trevPBE}{translated \acs{revPBE}}
	\acro{tBLYP}{translated \acl{BLYP}}
	\acro{ftSVWN3}{fully \acl{tSVWN3}}
	\acro{ftPBE}{fully \acl{tPBE}}
	\acro{ftBLYP}{fully \acl{tBLYP}}
	\acro{v2RDM}{variational \acl{2-RDM}}
	\acro{v2RDM-CASSCF}{\acl{v2RDM}-driven \acl{CASSCF}}
	\acro{v2RDM-CAS}{\acl{v2RDM}-driven \acl{CAS}}
	\acro{CAS-PDFT}{\acl{CAS} \acl{PDFT}}
	\acro{v2RDM-CASSCF-PDFT}{\acl{v2RDM} \acl{CASSCF} \acl{PDFT}}
	\acro{MAX}{maximum absolute error}
	\acro{CAM}{Coulomb-attenuating method}
	\acro{AKEE}{absolute kinetic energy error}
	\acro{MAKEE}[$\MAKEE$]{mean absolute kinetic energy error per electron}
	\acro{AKEEPE}[$\AKEEe$]{absolute kinetic energy error per electron}
	\acro{TMAE}[$\MAE$]{total \acl{MAE} per electron}
	\acro{MAEPE}[$\MAEe$]{\acl{MAE} per electron}
	\acro{NIAD}{normed integral absolute deviation}
	\acro{TNIAD}[$\NIAD$]{total normed integral absolute deviation}
	\acro{SNIAD}{spherical \acl{NIAD}}
	\acro{MAE}{mean absolute error}
	\acro{NPE}{non-parallelity error}
	\acro{HOMO}{highest-occupied \acl{MO}}
	\acro{LUMO}{lowest-unoccupied \acl{MO}}
	\acro{OEP}{optimized effective potential}
	\acro{LEB}{local energy balance}
	\acro{EKT}{extended Koopmans theorem}
\end{acronym}

	\bibliography{v2rdm_gh_pdft}
	
\end{document}